\documentstyle[preprint,aps]{revtex}
\begin{document}
\draft
\title{The scalar sector of 3-3-1 models}
\author{M. D. Tonasse\footnote{Departamento de F\'\i sica Nuclear e Altas
 Energias. E-mail: Tonasse@vax.fis.uerj.br.}}
\address{Universidade do Estado do Rio de Janeiro, Instituto de F\'\i sica,
 Rua S\~ao Francisco Xavier, 524, 20550-013 Rio de Janeiro, RJ, Brazil}
%\date{\today}
\maketitle

\begin{abstract}
We study the mass spectrum and the eigenstates of the scalar sectors in 
3-3-1 models. We show that, in one of the models, the physical scalar masses 
lead to theoretical constraints to the vacuum expectation values. The models 
allow very light Higgs bosons. One of the neutral scalars can be identified 
with the standard model one.
\end{abstract}

\vskip 1 true cm


\section{Introduction}
\label{intro}

The 3-3-1 models of electroweak interactions\cite{PP92}, based on the 
SU(3)$_L\otimes$U(1)$_N$ gauge group, predict some new particles which lead 
to new physics and interesting phenomenology. In particular, these models 
predict no very high new mass scales, in such a way that they can be confirmed 
or ruled out in a near future~\cite{FN92,AF92}. They belong to a type of gauge 
models which has the SU(2)$\otimes$U(1) electroweak doublet contained in a 
SU(3)$\otimes$U(1) triplet. The SU(3)$_L\otimes$U(1)$_N$ $\longmapsto$ 
SU(2)$_L\otimes$U(1)$_Y$ break scale is estimated by running 
$\sin^2{\theta_W}$ towards large values, which give the bound $w \lesssim 3$ 
TeV~\cite{NG94}.\par
This type of models, that includes the unified ones and chiral extensions of 
the standard model, is characterized by SU(2) doublets 
$\left(\begin{array}{cc} X^\pm & X^{\pm\pm}\end{array}\right)$ of vector and 
scalar gauge bosons which carry lepton number L = 2 (dileptons). The dileptons 
mediate processes which violate individual lepton numbers. The total lepton 
number ($L = L_e + L_\mu + L_\tau$) is conserved.\par
There are some aspects which turn the 3-3-1 models a very interesting 
extension of the standard model\cite{PP92,NG94,PP96}. The main two of them are 
the prediction about the family number (which must be multiple of the color 
number) and the bound for the Weinberg angle ($\sin{\theta_W} < 1/4$). 
Experimentally an interesting feature is that the mass scale of the vector 
dileptons can be as low as the weak scale. Thus, the lower bound on these 
masses can be obtained from the available data of the weak processes such as 
muon decay, low energy neutral current experiments, Bhabha scattering, 
etc\cite{AF92,CF92}.\par
In this work we discuss the Higgs potential of the 3-3-1 models at the tree 
level. The phenomenology of the scalar sector of the 3-3-1 models is not 
studied yet. In the model of Refs. \cite{PP92}, the minimal number of Higgs 
multiplets required in order to generate all fermion masses is four: three 
triplets and one sextet. The latter is necessary in order to get the masses of 
the charged leptons. However, if we assume the presence of heavy 
leptons\cite{PT93a} the sextet is not necessary. Also, if a right-handed 
singlet neutrino is added, it is possible to generate the right mass spectrum 
of the charged lepton through radiative corrections without introducing the 
sextet of Higgs bosons~\cite{PP93}. Thus, we can consider the VEV of the 
sextet rather small (a few GeV) relative to the VEV's of the triplets which 
are $\sim$ 100 GeV (two) and $\sim$ (1 -- 3) TeV (one). So, the three triplet 
models can be a good approximation for the scalar sector of the 3-3-1 
models.\par
Here we study both, the three triplet and the three triplet and one sextet 
models. Each of them has neutral CP-even and CP-odd scalar bosons as well as 
singly charged and doubly charged ones. In this type of models there can be 
charged scalar masses of the same order (or smaller) than the neutral scalar 
ones. In the approximation which we are using here it is always possible to 
identify one of the Higgs bosons with the scalar of the minimal 
SU(2)$_L \otimes$U(1)$_Y$ model.\par
\bigskip
In the next section we discuss the Higgs potentials of the models, presenting 
the masses and eigenstates for the two 3-3-1 models which are considered here. 
In Sec. 3 we summarize our results.

\section{The scalar Potentials}

The details of the 3-3-1 models were already pointed out in several works 
(see, for example, Refs. \cite{PP92,NG94}). Therefore, we write below only the 
scalar fields. The scalar sectors which are considered here are the minimal 
ones. We also do not consider CP violation through the scalar 
exchange\cite{LN94}.

\subsection{The Three Triplet Model}

In this case the scalar sector is composed by the three SU(3) triplets
\begin{equation}
\eta = \begin{array}{cccccc}
\left(\begin{array}{c}
\eta^0 \\ \eta^-_1 \\ \eta^+_2
\end{array}\right) \sim ({\bf3},0); \qquad
\rho = \left(\begin{array}{c}
\rho^+ \\ \rho^0 \\ \rho^{++}
\end{array}\right) \sim ({\bf3},1); \qquad
\chi = \left(\begin{array}{c}
\chi^- \\ \chi^{--} \\ \chi^0\end{array}\right) \sim ({\bf3},-1).
\end{array}
\label{tri}
\end{equation}
The most general, renormalizable and gauge invariant  
SU(3)$_L\otimes\-$U(1)$_N$ Higgs potential which we can write with the three 
triplets of the Eqs. (\ref{tri}) is given by
\begin{eqnarray}
V_T(\eta,\rho,\chi)& = &\mu_1^2\eta^\dagger\eta+\mu^2_2\rho^\dagger\rho +
\mu^2_3\chi^\dagger\chi+
\lambda_1(\eta^\dagger\eta)^2 + \lambda_2(\rho^\dagger\rho)^2 + 
\lambda_3(\chi^\dagger\chi)^2 
+\nonumber
\\ & &\mbox{}
+(\eta^\dagger\eta)\left[\lambda_4(\rho^\dagger\rho )+\lambda_5(\chi^\dagger
\chi)\right] +\lambda_6(\rho^\dagger\rho)(\chi^\dagger\chi) + 
\lambda_7(\rho^\dagger\eta)(\eta^\dagger\rho) \nonumber +\\ 
& &\mbox{} + \lambda_8(\chi^\dagger\eta)(\eta^\dagger\chi) + 
\lambda_9(\rho^\dagger\chi)(\chi^\dagger\rho) +
\left(\frac{f_1}{2}\varepsilon^{ijk}\eta_i\rho_j\chi_k + H. c.\right),
\label{potential1}
\end{eqnarray}
where the $\mu$'s, $\lambda$'s and $f_1$ are coupling constants, the latter 
with dimension of mass. Since the original model of Refs.\cite{PP92} do not 
consider neutrino mass\cite{PP94}, the potential (\ref{potential1}) is 
$F$-conserving\cite{PT93b}, with $F = L + B$, where $L$ is the total lepton 
number and $B$ is the baryon number.\par
Let the neutral scalars $\eta^0$, $\rho^0$, and $\chi^0$ acquire a vacuum 
expectation value $v_\eta$, $v_\rho$, and $v_\chi$, respectively, and define
\begin{equation}
\varphi = v_\varphi + \xi_\varphi + i\zeta_\varphi,
\label{desloc}
\end{equation}
with $\varphi = \eta^0$, $\rho^0$, $\chi^0$. We will use the notation 
$v_\eta \equiv v$, $v_\rho \equiv u$, and $v_\chi \equiv w$. The pattern of 
the symmetry breaking is\par
\medskip
\centerline{SU(3)$_L\otimes$ U(1)$_N$ 
$\stackrel{\langle\chi\rangle}{\longmapsto}$ SU(2)$_L\otimes$
U(1)$_Y$ $\stackrel{\langle\rho,\eta\rangle}{\longmapsto}$ U(1)$_{\rm em}$,}
\medskip\par
\noindent and the VEV's are related with the standard model one ($v_W$) as 
$v^2 + u^2 = v_W^2$.\par
The $\xi_\varphi$ $\left(\zeta_\varphi\right)$ fields lead to three CP-even 
(one CP-odd) physical scalar bosons. The exact form of the CP-even mass and 
eigenstates are troublesome. Thus, we will give the masses and eigenstates of 
this sector in an approximate form and for clearing this point we will 
consider this approximation in some detail. The square mass matrix coming from 
the $\xi_\varphi$ fields reads as
\begin{equation}
M_\xi^2 = \frac{1}{2}\left(
\begin{array}{cccc}
8\lambda_1v^2 - f_1uw/v & 4\lambda_4vu + f_1w & 4\lambda_5vw + f_1u \\ 
4\lambda_4vu + f_1w & 4\lambda_2u^2 - f_1vw/u & 4\lambda_6uw + f_1v \\ 
4\lambda_5vw + f_1u & 4\lambda_6uw + f_1v & 8\lambda_3w^2 - f_1vu/w
\end{array}
\right)
\label{cpparmass}
\end{equation}
in the $\xi_\eta$, $\xi_\rho$, $\xi_\chi$ basis. Here we have already imposed 
the constraints due to the non-linearity of the shifted potential in the 
$\xi_\varphi$ and $\zeta_\varphi$ fields and we have not considered phases in 
the VEV's. A natural approximation is to impose $\vert f_1\vert \sim w$ (this 
also avoids a new mass scale beyond $w$ in the model) and to maintain only 
terms of higher order in $w$ in Eq. (\ref{cpparmass}). But this results in one 
zero mass (H$^0_1$) and two massive physical states (H$^0_2$ and H$^0_3$) with 
masses 
\begin{equation}
m_{02}^2 \approx \frac{v^2 + u^2}{2vu}w^2, \qquad
m_{03}^2 \approx -4\lambda_3w^2,
\label{masseven2}
\end{equation}
associated with the approximate eigenstates
\begin{mathletters}
\begin{eqnarray}
\left(\begin{array}{cccc} \xi_\eta \\
                          \xi_\rho \end{array}\right) & \approx & 
\frac{1}{\left(v^2 + u^2\right)^{1/2}}
\left(\begin{array}{cccc} 
v \quad &  u \\
u \quad & -v 
\end{array}\right) 
\left(\begin{array}{cccc} H^0_1 \\
                          H^0_2 \end{array}\right), \\
\label{eigreal1}
\xi_\chi & \approx & H^0_3.
\label{eigreal2}
\end{eqnarray}
\end{mathletters}
Thus, the former eigenstate (H$^0_1$) must be a {\it light} Higgs boson, 
similar to that of the standard model. In order to confirm this assertion and 
to improve the approximation we take the eigenvalue equations with the exact 
mass matrix of Eq. (\ref{cpparmass}) for the eigenvalue $m_{01}^2$, associated 
with H$_1^0$ field, with the correspondent approximate eigenstate and identify 
a new contribution to the initially null $m_{01}^2$ eigenstate. So, it is easy 
to see that the H$^0_1$ is a state associated with the mass
\begin{equation}
m_{01}^2 \approx 4\frac{\lambda_2u^4 - \lambda_1v^4}{v^2 - u^2},
\label{masseven1}
\end{equation}
with the following relations among the coupling constants and VEV's:
\begin{equation}
\lambda_4 \approx 2\frac{\lambda_2u^2 - \lambda_1v^2}{v^2 - u^2}, \qquad 
\lambda_5v^2 + 2\lambda_6u^2 \approx -\frac{vu}{2}.
\end{equation}
Notice that the $m_{01}^2$ mass of Eq. (\ref{masseven1}) is a function of $v$ 
and $u$ VEV's only. So, this is the {\it light} scalar which we identify with 
the standard model Higgs boson. However, since the $\lambda$'s are still 
unbounded and the {\it light} Higgs mass is function of $\left(v^2 - 
u^2\right)^{-1}$, we have a large range for the mass scales, as in the 
standard model.\par
Concerning the imaginary part of the right-hand side of the Eq. (\ref{desloc}) 
($\zeta_\varphi$ fields) we have two Goldstone bosons and one massive physical 
state with exact mass
\begin{equation}
m^2_h = -\frac{f_1w}{vu}\left[v^2 + u^2 + \left(\frac{vu}{w}\right)^2\right],
\label{massodd}
\end{equation}
where $f_1 < 0$.\par
The symmetry eigenstates in the CP-odd sector are related to the physical 
states as follows:
\begin{mathletters}
\begin{eqnarray}
\left(\begin{array}{cccc} \zeta_\eta \\
                          \zeta_\rho \end{array}\right) & \approx &
-\frac{1}{\left(v^2 + u^2\right)^{1/2}}
\left(\begin{array}{cccc}
                 v \quad & u \\
                 u \quad & v \end{array}\right)
\left(\begin{array}{cccc} G^0_1 \\
                          G^0_2 \end{array}\right), \\
\zeta_\chi & \approx & h^0,
\end{eqnarray}
\end{mathletters}
where the G$^0$'s states are neutral Goldstone bosons and h$^0$ is the CP-odd 
massive neutral physical state.\par 
In the charged scalar sector, we have obtained all masses and eigenstates 
exactly. In the singly charged sector we have two Goldstone bosons and two 
physical Higgs whose masses read
\begin{equation}
m^2_{+1} = \frac{v^2 + u^2}{2vu}\left(f_1w - 2\lambda_7vu\right), \qquad 
m^2_{+2} = \frac{v^2 + w^2}{2vw}\left(f_1u - 2\lambda_8vw\right),
\label{massh+}
\end{equation}
with mixings
\begin{mathletters}
\begin{eqnarray}
\left(\begin{array}{cccc} \eta^+_1 \\
                          \rho^+ \end{array}\right) & = &
\frac{1}{\left(v^2 + u^2\right)^{1/2}}\left(\begin{array}{cccc} -v & u \\
                                                     u & v \end{array}\right)
\left(\begin{array}{cccc} G^+_1 \\
                          H^+_1 \end{array}\right),
\label{eig1+} \\
\left(\begin{array}{cccc} \eta_2^+ \\
                          \chi^+ \end{array}\right) & = & 
\frac{1}{\left(v^2 + w^2\right)^{1/2}}\left(\begin{array}{cccc} -v & w \\
                                                     w & v \end{array}\right)
\left(\begin{array}{cccc} G_2^+ \\
                          H^+_2 \end{array}\right).
\label{eig2+}
\end{eqnarray}
\end{mathletters}
\par
In the doubly charged sector there is only one doubly charged Goldstone and 
one physical Higgs boson with mass
\begin{equation}
m^2_{++} = \frac{u^2 + w^2}{2uw}\left(f_1v - 2\lambda_9uw\right).
\label{massh++}
\end{equation}
The doubly charged mass eigenstates are given by
\begin{equation}
\left(\begin{array}{cccc} \rho^{++} \\
                          \chi^{++} \end{array}\right) =
\frac{1}{\left(u^2 + w^2\right)^{1/2}}\left(\begin{array}{cccc} -u & w \\
                                                     w & u \end{array}\right)
\left(\begin{array}{cccc} G^{++} \\
                          H^{++} \end{array}\right).
%\label{eig++}
\end{equation}\par
From Eqs. (\ref{masseven2}), (\ref{masseven1}), (\ref{massodd}), 
(\ref{massh+}) and (\ref{massh++}) we have the conditions for the 
potential to be below bounded:
\begin{mathletters}
\begin{eqnarray}
\frac{\lambda_1}{\lambda_2} & & \left\{ 
          \begin{array}{rl} \lesssim u^4/v^4, & {\rm if} \quad v > u; \\
                            \stackrel{>}{\sim} u^4/v^4, & {\rm if} \quad v < u;
           \end{array}\right.
\label{cond1} \\
v \neq u, \quad \lambda_3 \lesssim 0, \quad f_1 < 0; \qquad 
\frac{f_1}{\lambda_7} & < & 2\frac{vu}{w}, \quad \frac{f_1}{\lambda_8} < 
2\frac{vw}{u}, \quad 
\frac{f_1}{\lambda_9} < 2\frac{uw}{v}.
\label{cond2}
\end{eqnarray}
\end{mathletters}
We recall that the masses of the CP-odd neutral scalar and the charged ones 
are exact.\par
The {\it light} Higgs boson H$^0_1$ can be of interest since it {\it imitates} 
the standard model one. The trilinear (H$_1^0$Z$^0$Z$^0$) and quartic 
(H$_1^0$H$_1^0$Z$^0$Z$^0$) interactions of the {\it light} neutral scalar 
boson are
\begin{mathletters}
\begin{eqnarray}
&& g^2\frac{\sqrt{3}\left[u^2w^2 + 3\left(u^4 + 2v^2w^2 + 
2w^4t^4\right)\right] + 6uw\left(v^2 + w^2t^2\right)}{6\sqrt{3}u^2w^2}, \\
&& g^2\frac{\sqrt{3}\left(u^2w^2 + 3v^4\right) + 3w\left[\left(v^2 + 
t^2w^2\right)\left(2u + \sqrt{3}t^2w\right) + 
2\sqrt{3}u^2wt\right]}{12\sqrt{3}v^2u^2},
\end{eqnarray}
\end{mathletters}
where $t \equiv g^\prime/g$ is the ratio among the coupling constants of the 
U(1) and the SU(3) groups, respectively. Since the coupling strengths are very 
different from the standard model ones (and other popular gauge models), the 
current results of the collider experiments on mass bounds are not directly 
applicable to 3-3-1 scalar bosons.\par
The scalar masses are functions of the unknown scalar potential parameters and 
of the VEV's of the scalar neutral fields which are poorly bounded and so, the 
range for scalar mass scale is large.\par
As we comment in the Introduction the VEV of the sextet neutral field must be 
small in such a way that the three triplet case be a good approximation to the 
original model of Ref. \cite{PP92}. If we also assume the condition $w \sim 
-f_1$ in the charged sector, the masses of the singly charged scalar [Eqs. (
\ref{massh+})] and the doubly charged [Eq. (\ref{massh++})] scale with $w^2$, 
like the neutral scalar H$^0_2$ in Eqs. (\ref{masseven2}). In particular we 
have in this approximation $m_{02} = m_{+2}$. On the other hand, $m_{01}$ in 
Eqs. (\ref{masseven1}) can be large for some set of parameters. Therefore, it 
is possible in this model a charged Higgs mass be smaller than a neutral one. 
If the decay of the neutral Higgs in pairs of charged ones is kinematically 
allowed it is a good channel for Higgs searching\cite{DL79}. The H$_0^1$ 
scalar boson is similar to the standard model one. The similarity becomes 
exact if we put $v$ $\left(u\right) = 0 \left[\mbox{with } v^2 + u^2 = 
\left(246 \mbox{ GeV}\right)^2\right]$. In such cases the SU(2) scalar doublet 
is embedding in the $\eta$ $\left(\rho\right)$ 3-3-1 triplet of Eqs. 
(\ref{tri}).\par
A graphical analysis of $m_{01}$ and the charged scalar masses is spoiled by 
our ignorance on the Higgs potential constants. However, we plot in the Figure 
1 the curves for $m_{02}$ and $m_h$, which show that these masses can be 
relatively small for a large range of $v$ VEV. There is pratically no 
difference for 1 TeV $< w <$ 3 TeV.\par

\subsection{The Three Triplet and One Sextet Model}

Here we need to add the sextet
\begin{equation}
S=\left(\begin{array}{ccc}
\sigma^0_1 & s_2^+ & s_1^- \\
s_2^+ & S_1^{++} & \sigma_2^0 \\
s_1^- &\sigma_2^0 & S_2^{--} 
\end{array}\right) \sim ({\bf 6}^*, 0)
\label{sextet}
\end{equation}
of scalar fields to the three triplets of Eqs. (\ref{tri}). Thus we have 
additional terms in the Higgs potential of Eq. (\ref{potential1}). The new 
potential is
\begin{eqnarray}
V_S\left(\eta , \rho , \chi, S\right) & = & V_T + \mu^2_4
\mbox{Tr}\left(S^\dagger 
S\right) + \lambda_{10}\mbox{Tr}^2\left(S^\dagger S\right) + 
\lambda_{11}\mbox{Tr}\left[\left(S^\dagger S\right)^2\right] + \cr
&& + \left[\lambda_{12}\left(\eta^\dagger\eta\right) + 
\lambda_{13}\left(\rho^\dagger\rho\right) + 
\lambda_{14}\left(\chi^\dagger\chi\right)\right]
\mbox{Tr}\left(S^\dagger S\right) + \cr
&& + \frac{1}{2}\left(f_2\rho_i\chi_jS^{ij} + \mbox{H.c.}\right),
\label{potential2}
\end{eqnarray}
where $V_T$ is given in Eq. (\ref{potential1}). The VEV of the $\sigma^0_2$ 
field is $v^\prime$ and that for $\sigma^0_1$ vanishes, since we do not 
consider neutrino mass\cite{PP92}. Eq. (\ref{desloc}) have now an additional 
component given by $\sigma^0_2 = v^\prime + \xi^\prime + i\zeta^\prime$. The 
$v^\prime$ VEV contributes to the SU(2)$_L\otimes$U(1)$_N$ $\longmapsto$ U
(1)$_{\rm em}$ breaking in such a way that $v^2 + u^2 + {v^\prime}^2 = 
v_W^2$.\par
Working with the same approximation as in the three triplet case (we are c
onsidering also $\vert f_2\vert \sim w$), we obtain the following square 
masses from the neutral CP-even sector:
\begin{mathletters}
\begin{eqnarray}
M^2_{01} & \approx & -4\lambda\left(v^2 + u^2 + {v^\prime}^2\right), 
\label{leve}\\
M^2_{02,03} & \approx & \frac{w^2}{4vuv^\prime}
\left\{\left(su^2 + rvv^\prime\right)w + \right. \cr
& &\left. \pm \left[\left(su^2 - rvv^\prime\right)^2w^2 - 
\frac{4vu^2v^\prime}{w^2}\left[f_1f_2u^2 - \left(f_1^2 + 
f_2^2\right)vv^\prime\right]\right]^{1/2}\right\}, \\
M^2_{04} & \approx & -4\lambda_3w^2.
\end{eqnarray}
\end{mathletters}
where we are defining 
\begin{mathletters}
\begin{eqnarray}
r & \equiv & \frac{1}{w^2}\left(f_1v - wv^\prime\right) \approx 
-\frac{2}{uw}\left(\lambda_5v^2 + \lambda_6u^2 + 
2\lambda_{14}{v^\prime}^2\right), \\
s & \equiv & \frac{1}{w^2}\left(f_1v^\prime - vw\right).
\end{eqnarray}
\end{mathletters}
This approximation implies
\begin{equation}
\lambda \equiv 4\left(2\lambda_{10} + \lambda_{11}\right) \approx \lambda_1 
\approx \lambda_2 \approx \lambda_{12} \approx \lambda_{13} \approx 
\frac{\lambda_4}{2}.
\end{equation}
In this case the standard model Higgs boson corresponds to H$_1^0$, in the 
sense that its mass has no dependence in the $w$ parameter of the symmetry 
breaking [see Eq. (\ref{leve})].\par
Defining
\begin{mathletters}
\begin{eqnarray}
N_\rho^2 & \equiv & \frac{1}{w^4}\left\{\left[u\left(uw^2 - 
2M^2_{04}v^\prime\right) + w^2{v^\prime}^2\right]^2 - 
vw\left[\left(uw^2 + 2M^2_{04}v^\prime\right)^2 + 
w^4{v^\prime}^2\right]\right\}, \\
N_\chi^2 & \equiv & \frac{1}{w^{12}}\left\{\left(v^2 + 
{v^\prime}^2\right)\left[v\left(2M_{04}^2v^\prime + uw^2\right) + 
uw^3\right]^2w^4 + \right. \cr
& & \left. + \left[u\left(uw^2 + 2M_{04}^2v^\prime\right) + 
\left({v^\prime}^2 + vw\right)w^2\right]^2\times \right. \cr
& & \left. \times\left[\left(u^2 + {v^\prime}^2\right)w^4 + 
2vu\left[v\left(2M_{04}^2v^\prime + uw^2\right) + uw^3\right]\right]\right\}, \\
a_{21} & \equiv & \left[u\left(uw^2 + 2M_{04}^2v^\prime\right) + 
v^\prime w^2\right]/w^2N_\rho, \\
a_{22} & \equiv & -v\left(2M_{04}^2v^\prime + uw^2\right)/w^2N_\rho, \\
a_{31} & \equiv & v^\prime\left[-v\left(2M_{04}^2v^\prime + uw^2\right) + 
uw^3\right]/w^5N_\chi, \\
a_{32} & \equiv & v^\prime\left[u\left(uw^2 + 2M_{04}^2v^\prime\right) + 
\left({v^\prime}^2 + vw\right)w^2\right]/w^5N_{\chi}, \\
a_{33} & \equiv & \left[-\left(2M_{04}^2v^\prime + 
uw^2\right)\left(v^2 - u^2\right) + u{v^\prime}^2\right]/w^4N_{\chi}
\end{eqnarray}
\end{mathletters}
we have the eigenstates 
\begin{mathletters}
\begin{eqnarray}
\left(\begin{array}{cccc}
      \xi_\eta \\
      \xi_\rho \\
      \xi_\chi
\end{array}\right) & \approx & 
\left(\begin{array}{cccc}
2\sqrt{\vert\lambda\vert}v/M_{01} & 2\sqrt{\vert\lambda\vert}u/M_{01} & 
2\sqrt{\vert\lambda\vert}v^\prime/M_{01} \\ 
a_{21} & a_{22} & vv^\prime/N_\rho \\
a_{31} & a_{32} & a_{33}
\end{array}\right)
\left(\begin{array}{cccc}
   {\cal H}^0_1 \\
   {\cal H}^0_2 \\
   {\cal H}^0_3
\end{array}\right),
\label{eigven1} \\
\xi^\prime & \approx & {\cal H}_4^0.
\label{eigven2}
\end{eqnarray}
\end{mathletters}
In the CP-odd part we have two Goldstone and two physical Higgs bosons with 
masses:
\begin{eqnarray}
M^2_{{\cal J}1,{\cal J}2} & \approx & 
\frac{w^2}{4vuv^\prime}\left\{\left(su^2 + rvv^\prime\right)w + \right. \cr
& & \left. \pm \left[\left(su^2 - rvv^\prime\right)^2w^2 - 
\frac{4vu^2v^\prime}{w^2}\left(f_1f_2u^2 + \left(f_1^2 - 
f_2^2\right)vv^\prime\right)\right]^{1/2}\right\}.
\end{eqnarray}
In the same way, for the CP-odd sector, the mass eigenstates read
\begin{mathletters}
\begin{eqnarray}
\left(\begin{array}{cccc}
      \zeta_\eta \\
      \zeta_\rho \\
      \zeta_\chi
\end{array}\right) & \approx & 
\left(\begin{array}{cccc}
2\sqrt{\vert\lambda\vert}v/M_{01} & -2\sqrt{\vert\lambda\vert}u/M_{01} & 
2\sqrt{\vert\lambda\vert}v^\prime/M_{01} \\ 
b_{21} & b_{22} & b_{23} \\
u/\left(v^2 + u^2\right)^{1/2} & v/\left(v^2 + u^2\right)^{1/2} & 0
\end{array}\right)
\left(\begin{array}{cccc}
   {\cal G}^0_1 \\
   {\cal G}^0_2 \\
   {\cal J}^0_1
\end{array}\right),
\label{eigodd1} \\
\zeta^\prime & \approx & {\cal J}_2^0,
\label{eigodd2}
\end{eqnarray}
\end{mathletters}
where
\begin{equation}
b_{21} \equiv \frac{2\sqrt{\vert\lambda\vert} vv^\prime}{M_{01}\left(v^2 + 
u^2\right)^{1/2}}, \qquad
b_{22} \equiv -\frac{2\sqrt{\vert\lambda\vert} uv^\prime}{M_{01}\left(v^2 + 
u^2\right)^{1/2}}, \qquad 
b_{23} \equiv \frac{2\left[\sqrt{\vert\lambda\vert}\left(v^2 + 
u^2\right)\right]^{1/2}}{M_{01}}.
\end{equation}\par
In this approximation two of the masses of the singly charged scalar bosons 
coincide with the CP-odd ones $\left[M_{+1} \, \left(M_{+2}\right) = M_{{\cal 
J}1} \, \left(M_{{\cal J}2}\right) \right]$. The other two are degenerated:
\begin {equation}
M^2_{+3} = M^2_{+4} \approx f_2\frac{uw}{2v^\prime},
\end{equation}
where $f_2 > 0$. The associated eigenstates are
\begin{mathletters}
\begin{eqnarray}
\left(\begin{array}{cccccc}
      \eta_1^+ \\
        \rho^+ \\
      \eta_2^+ 
\end{array}\right) & \approx & 
\left(\begin{array}{cccccc}
2\sqrt{\vert\lambda\vert}v/M_{01} & -2\sqrt{\vert\lambda\vert}u/M_{01} & 
-2\sqrt{\vert\lambda\vert}v^\prime/M_{01} \\ 
2\sqrt{\vert\lambda\vert}w/M_{01} & 
2\sqrt{\vert\lambda\vert}vu/\left[M_{01}\left(v^2 + u^2\right)\right] & a_{21} 
\\
0 & v^\prime/\left(u^2 + {v^\prime}^2\right)^{1/2} & -u/\left(u^2 + 
{v^\prime}^2\right)^{1/2}
\end{array}\right)
\left(\begin{array}{cccccc}
   {\cal G}^+_1 \\
   {\cal G}^+_2 \\
   {\cal H}^+_1
\end{array}\right),
\label{eig+1} \\
\chi^+ & \approx & {\cal H}_2^+, \qquad s_1^+ \approx {\cal H}_3^+, \qquad 
s_2^+ \approx {\cal H}_4^+.
\label{eig+2}
\end{eqnarray}
\end{mathletters}
Finally, for the doubly charged scalars the masses are
\begin{mathletters}
\begin{eqnarray}
M^2_{++1,++2} & \approx & \frac{w^2}{4v^\prime}\left\{u -2\lambda_9v^\prime 
\pm \left[\left(2\lambda_9v^\prime + u\right)^2 + 
4{v^\prime}^2\right]^{1/2}\right\}, \\
M^2_{++3} & \approx & f_2\frac{uw}{2v^\prime}
\end{eqnarray}
\end{mathletters}
Notice that $M_{+3}^2 = M_{+4}^2 = M^2_{++3}$ in this approximation. Here we 
have the eigenstates
\begin{mathletters}
\begin{eqnarray}
\left(\begin{array}{cccc}
      \rho^{++} \\
      \chi^{++}
\end{array}\right) & \approx & 
\frac{1}{\left(u^2 + {v^\prime}^2\right)}\left(\begin{array}{cccc}
u & v^\prime \\ 
v^\prime & -u
\end{array}\right)
\left(\begin{array}{cccc}
   {\cal G}^{++} \\
   {\cal H}^{++}_1
\end{array}\right),
\label{eig++1} \\
S_1^{++} & \approx & {\cal H}_2^{++}, \qquad S_2^{++} \approx {\cal H}_3^{++}.
\label{eig++2}
\end{eqnarray}
\end{mathletters}
We show the behavior of the $M_{02}$ and $M_{03}$ masses as functions of $v$ 
and $u$ VEV's in surface graphics in Figure 2. The physical values for 
$M_{03}$ constraint the $u$ VEV in range 178 GeV $\lesssim$ $u$ $<$ 246 GeV 
(Figure 2b). This mass increases indefinitely as $v \longmapsto 0$ and $u 
\longmapsto 246$ GeV. The behavior of the CP-odd neutral masses and singly 
charged ones are given in surface graphics in the Figure 3. We can see that 
the $v$ VEV is constrained in these figures by 0 $<$ $v$ $\lesssim$ 37 GeV. 
The behavior of the $M_{02} \left(M_{03}\right)$ is similar to $M_{{\cal J}1} 
\left(M_{{\cal J}2}\right)$. We study the doubly charged scalar masses as 
functions of the ratio $u/v^\prime$ and $\lambda_9$. $M_{++1}$ is small when 
$u/v^\prime$ and $\lambda_9$ constant are near zero and increases when these 
variables leave from the origin. We do not plot the correspondent graphic. 
$M_{++2}$ has a more interesting behavior. In Figure 4 we plot the curve for 
$M_{++2}$ as function of $u/v^\prime$ for $\lambda_9 = -0.5$ and $-1$. These 
curves show that $M_{++2}$ has a limiting value as $u/v^\prime$ increases. 
This upper value increases as $\lambda_9$ decreases.
 
\section{Summary}

Summarizing, we have studied the scalar sector of the 3-3-1 models in the {\it 
natural} condition $v, u, v^\prime \ll w$. It is showed that the 3-3-1 models 
have neutral and charged scalar masses which can be small for a large range of 
the parameters, turning the scalar sector of these models interesting for 
searching of Higgs bosons in available and future colliders. In 3-3-1 models 
is hopeful to find neutral and charged scalar Higgs bosons with masses which 
are not heavier than a few TeVs. One of the doubly charged Higgs bosons can be 
relatively very light and so, as in two doublets models\cite{HM95}, it can be 
used as a probe of the 3-3-1 models. It is interesting to observe here two 
aspects: firstly, in our approach the scalar masses do not depend on many free 
parameters and secondly, in the model with sextet, we are able to obtain {\it 
natural} constraints on the two VEV's ($0 < v \lesssim 40$ GeV and 180 GeV 
$\lesssim u <$ 246 GeV) independently of the experimental results.\par
Notice that the model with sextet is more predictive. Our analysis in this 
case would be improved if we had constraints for small $v$ VEV. This might be 
obtained through experimental data. We cannot put bound on the 3-3-1 
parameters by requiring the validity of the perturbative regime, as it is done 
in two Higgs doublets models\cite{BH90}. In our case the quark--Higgs coupling 
structure and the presence of new quarks\cite{PP92} with unknown masses spoil 
these bounds.

\acknowledgments

I thank the hospitality of Prof. C. O. Escobar and Prof. R. Zukanovich Funchal 
at the {\it Instituto de F\'\i sica} -- USP, where part of this work was done 
and to my friends from the {\it Instituto de F\'\i sica Te\'orica} -- UNESP, 
specially to Prof. V. Pleitez for helpful discussions about this work. I also 
thank Dr. M. C. Tijero for reading the manuscript and the Brazilian Agencies 
FAPESP ({\it Funda\c c\~ao de Amparo \`a Pesquisa no Estado de S\~ao Paulo}) 
and CNPq ({\it Conselho Nacional de Desenvolvimento Cient\'\i fico e 
Tecnol\'ogico}) for financial support during part of the realization of this 
work.

\newpage
\centerline{FIGURE CAPTIONS}
\vskip 1 true cm

\noindent Figure 1. Graphics for the neutral Higgs masses $m_{02}$ (filled 
line) and for $m_h$ (dashed line), with $w$ = 3 TeV, in the tree triplet Higgs 
model. $m/w$ stands for $m_{02}/w$ or $m_h/w$.\par
\bigskip
\noindent Figure 2. A surface graphic shows the range for physical neutral 
$M_{02}/w$ (a) and $M_{03}/w$ (b) masses as functions of the VEV's $v$ and $u$ 
in the three triplet and one sextet model.\par
\bigskip
\noindent Figure 3. The same as in the figure 2, but $M/w$ in the third 
coordinate is $M_{{\cal J}1}/w$ or $M_{+1}/w$ (a) and $M_{{\cal J}2}/w$ or 
$M_{+2}/w$ (b).\par
\bigskip
\noindent Figure 4. The behavior of the $M_{++2}/w$ as function of the ratio 
$u/v^\prime$ of the VEV's for $\lambda_9$ = -0.5 (dashed line) and $\lambda_9$ 
= -1 (filled line).
\end{document}